# Theory of Polaron Resonance in Quantum Dots and Quantum-Dot Molecules

K.-M. Hung[*]

Dept. of Electronics Engineering, National Kaohsiung University of Applied Sciences, Kaohsiung, Taiwan 807, ROC.

**Abstract**

The theory of exciton coupling to photons and LO phonons in quantum dots (QDs) and quantum-dot molecules (QDMs) is presented. Resonant-round trips of the exciton between the ground (bright) and excited (dark or bright) states mediated by the LO-phonon alter the decay time and yield the Rabi oscillation. The initial distributions of the population in the ground and the excited states dominate the oscillating amplitude and frequency. This property provides a detectable signature to the information stored in a qubit made from QD or QDM for a wide range of temperature T. Our results presented herein provide an explanation to the anomaly on T-dependent decay in self-assembled InGaAs/GaAs QDMs recently reported by experiment.

## I. Introduction

Charge carriers that move in semiconductor QDs and QDMs provide a larger transition-dipole moment than atomic and molecular systems owing to the interaction with solid-state matter and a spatial variation of the band edge in QDs and QDMs, making applications in quantum information processing [1] and logical operation [2,3] feasible. In such applications, coherent manipulation of the excitonic wavefunction in QDs and QDMs at finite temperature is essential. A longer dephasing time (~1ns in self-assembled InGaAs/GaAs QDs [4-6] and QDMs [6,7]) than the manipulation time (~1ps [8]) is very important, because the coherence of the excitonic transition, or quantum computation, can not be retained in the scope that the dephasing time is comparable to or smaller than the manipulation time.

In QD and QDM systems, the carrier dephasing can be categorized into two parts: (1) the dephasing of spatial wavefunction of exciton and (2) the dephasing of the internal degrees of freedom of exciton, such as degenerate spin states. The former dephasing mainly attributed to, for examples, photon and real phonon scatterings, so called the excitonic decay, because the exciton can not incoherently reside in a spatial-confined state. The second dephasing relaxes the internal degrees of freedom of exciton from one of its degenerate states to the others or changes the phase relation of the superposition state composed of the degenerate states, while preserves the spatial coherence, i.e., conserves the number of excitons, so called the pure dephasing. In fact, the pure dephasing could also affect the decay time through, for example, the relaxation between spin-dark and spin-bright states according to the selection rule of photon emission [6]. Recently, both experimental and theoretical reports [9,10] have revealed that the virtual-phonon processes for both acoustic and optical phonons result to the second dephasing and exhibit a non-monotonous temperature dependence on rapid initial decoherence (~1ps) [11].

In principle, the charge cancellation for the identical distributions of electron and hole in a strong confining QD reduces the interaction of excitons with LO phonons [12-14], and a large level spacing decreases the strength of exciton scattering from real acoustic phonons. Accordingly, a long decay time is expected. However, the presence of piezoelectric fields [15], QD's shape and/or size fluctuations [16], the Jahn-Teller effect [17], and charged point defects [18] lead to polarization of the charge distributions, and thus enhance the LO phonon-exciton coupling. As for the electronic polaron [19], the coherent interactions of the electron-hole pair with a polarizable field (a real optical phonon) form excitonic polaron and do not contribute to phase decoherence, because the dressed state is an eigenvector of the interacting exciton-LO-phonon system. It has been shown experimentally that the resonance of excitonic polaron exists as the energy separation between the electronic states differs by one or several LO-phonons [20]. This could occur deficiently in bulk, quantum well, or quantum wire structures according to the phase decoherence through the process of real acoustic phonon, which is strongly suppressed in QDs and QDMs due to a large level spacing.

The decay may result from the photon emission, the coupling of the phonon thermostat that is originated by the anharmonicity of the crystal through the LO-phonon component [19], the thermal emission of carriers out of the dots at high temperature T [6], and the virtual phonon processes [9-11]. With neglecting the second effect, the delta-like density of states of QD prohibited k-space thermalization results in a nearly constant decay time with respect to T during the situations that thermal emission can be ignored and the time scale is far away from that of the pure dephasing, because the photon emission is independent of T. Yet, this feature differs dramatically in self-assembled InGaAs/GaAs QDs [21] and QDMs [6], for which the decay time increases with T. This feature cannot be described as the thermal recycling of the carriers [6,22] because it almost disappears in the InGaAs/GaAs QDs with the same structure as used in the QDMs [6]. Instead, the thermal population of optically inactive states gives a reasonable explanation on the increase of the decay time as increased the lattice temperature.

In this work, a theory of the coupling of excitons to photons and LO phonons associated with the dark or weak-bright excited states is proposed to explain the anomaly. Our theoretical results also reveal the time-dependent Rabi oscillation (RO) for both QD and QDM systems [23]. The oscillating frequency and the oscillating amplitude of the ROs are strongly dependent on the initial distributions of the excitonic population, which depend on the ways of excitation, shown in Fig. 1(a). This property provides a detectable signature to the information stored in a qubit made from QD or QDM with a long lifetime and a wide range of temperature.

In the following discussions, a model and its corresponding Hamiltonian are proposed in Sec. II. Based on the model Hamiltonian, the dynamics of exciton population in QDs/QDMs are derived in Sec. III. The numerical results and discussions for T-dependent decay time of an exciton in QD and QDM systems are presented in Sec. IV. Finally, a conclusion is given in Sec. V.

## II. Model and Hamiltonian

**II-A. Physical Picture and Model**

To describe the basic concept of the proposed theory, consider a three-level system of a QD or QDM embedded in a phonon bath as schematically plotted in Fig. 1(b). The ground (bright) state |g> of exciton is coupled to its vacuum state |vac> via the processes of emission/absorption of a photon, while coupled to the lowest excited state |e> by LO-phonon emission/absorption. The excited state can be a dark (or a weak bright) state in QDMs due to symmetric and anti-symmetric splitting for a symmetric structure [3] (or due to the inter-dot exciton for an asymmetric one) but a bright state in QDs, shown in Fig. 1(c). The same situation has been proposed in the study of exciton-enhanced Raman scattering in bulk semiconductors [24].

In order to explain how the brightness of the excited state associated with the phonon-assisted transition affect the decay time of an exciton in QDs/QDMs, let us imagine that there is an hourglass with a controlled gate at the vent to control the rate of the sand flow (analogous to the optical-transition rate of exciton). There are tow positions for the hourglass one at the upstairs (state |e>) and another at downstairs (state |g>). The gate is assumed to be position dependent, which is open (a bright exciton) at the downstairs with the flow (optical transition) rate $\Gamma_g$, but is open (bright exciton) or close (dark or weak-bright exciton) at the upstairs with the rate $\Gamma_e$ that depends on the system one considered. The hourglass could move up and down between the downstairs and upstairs (analogous to the RO of exciton). The dynamics of the hourglass depends on the external force (temperature) that forces the hourglass to move up and down between the downstairs and upstairs, and on its transition rate $\gamma_0$ (the total coupling strength between exciton and LO-phonon). When the hourglass statically stays in the downstairs (the absence of phonon-assisted transition at low T), the discharged time of the sand (the decay time of the exciton) is solely determined by the flow rate $\Gamma_g$ [25]. When the hourglass moves up and down between the upstairs and downstairs (the presence of RO), the discharged time will be governed by both the flow rates of the upstairs and downstairs and the dynamical distribution of the hourglass. In the (high T) case that the hourglass has a half time stayed in the downstairs and another half in the upstairs with an equal probability to move up and down, the averaged flow rate becomes $(\Gamma_g+\Gamma_e)/2$, and becomes $\Gamma_g/2$ or double the discharged (decay) time for $\Gamma_e=0$ or $\Gamma_e<<\Gamma_g$ (a fully-dark or weak-bright excited state). Hence, a fully dark-excited state results to the maximum enhancement of the decay time by a factor of two. In moderate temperature, the averaged rate has the form of $\sim N_{gNRO}(0^+)\Gamma_g + N_{eNRO}(0^+)\Gamma_e$, where $N_{gNRO}$ ($N_{eNRO}$) is the initial number of exciton in the state |g⟩ (|e⟩) with removing RO.

Although this model is simple, it sufficiently describes the mechanism of exciton decay caused by the interaction of exciton and LO-phonon. Spontaneous emission of a LO-phonon by an excited exciton, which is absent in the case that the exciton initially resides in its ground state, makes the T-dependent behavior of exciton very sensitive to its initial population distribution.

**II-B. Sudden Approximation**

In experiments, there are two excitation approaches, schematically plotted in Fig. 1(a), provided with a low excitation power for preventing charged exciton or biexciton effects often used in photoluminescence (PL) experiments: (i) indirect excitation — the excitons are initially created on the substrate or wetting layer and then relaxed into the states of QDs or QDMs. In such an excitation, a nonzero distribution of the exciton population in both the states |g⟩ and |e⟩ is possible. (ii) Direct excitation — the distribution of excitonic population is initially determined by the frequency of the excitation light source. The rise time of the number of excitons should depend on the mechanisms of carrier relaxation, optical transition, excitation power, and duration of excitation pulse.

The discussion of the rise time that is, in the most cases, much shorter than the decay time is beyond the scope of this work. To avoid the complicated processes of exciton generation and rapid initial decay we assume that the system could promptly response to an ultra-short pulse excitation, a sudden approximation. In the approximation, the exciton is abruptly present in the system at the time $0^+$, before that time all of the transitions caused by the interactions of exciton-phonon and exciton-photon are absent [26]. Although the sudden approximation can not occur in a real system, but it is a good approximation to the case that the decay time is much longer than the rise time, such as the exciton decay in InAs/GaAs QDs/QDMs [6], and is useful in the simplification of theoretical derivation.

**II-C. Hamiltonian**

In bosonized approximation, the interacting Hamiltonian of the exciton coupling to photons and LO-phonons (Fröhlich type) in a three-level system can be expressed as,

$$H_I = \sum_p (\gamma_{pg} \hat{a}_p^+ \hat{\alpha}_g + c.c.) + \sum_p (\gamma_{pe} \hat{a}_p^+ \hat{\alpha}_e + c.c.) + \sum_q (\gamma_q \hat{d}_q^+ \hat{\alpha}_g^+ \hat{\alpha}_e + c.c.),  \quad (1)$$

where $\hat{a}_p^+$ ($\hat{a}_p$), $\hat{d}_q^+$ ($\hat{d}_q$), and $\hat{\alpha}_i^+$ ($\hat{\alpha}_i$) are the creation (annihilation) operators of the photon, the LO phonon, and the exciton, respectively, $\gamma_q$ ($\gamma_{pi}$) is the strength of coupling of the exciton to the LO phonons (photons). A rotation-less approximation to the photon field is made, because the transition between spin-bright and spin-dark is not a major concern in this work. Notably, the exciton operators utilized in the pairing theory of superconductors [27] satisfy the algebra

$$[\hat{\alpha}_i, \hat{\alpha}_j^+] = \delta_{i,j}(1 - \hat{n}_{ei} - \hat{n}_{hi}) \quad (2)$$

because of the Pauli principle, where $\hat{n}_{e(h)i}$ is the number operator of electron (hole) in the state i. Although Eq. (2) is not in the form required by Bose-Einstein statistics, the factor $\hat{n}_{ei} + \hat{n}_{hi}$ does not affect the physics of the exciton's dynamics in counting all orders of resonant LO-phonon (conserved the phonon's momentum in the exciton loop as shown in Fig. 1(b), detailed in Sec. III-E) and in the second-order approximation to the photon correction. The definition of the exciton operator α_i, a product of the annihilation operators of the electron ($\hat{c}_i$) and the hole ($\hat{h}_i$) in the state i, gives the useful identities in the derivations of the equations of motion (EOM) of the exciton,

$$\hat{\alpha}_i \hat{\alpha}_i = \hat{\alpha}_i^+ \hat{\alpha}_i^+ = \hat{\alpha}_i^+ \hat{n}_{ei} = \hat{\alpha}_i^+ \hat{n}_{hi} = \hat{n}_{ei} \hat{\alpha}_i = \hat{n}_{hi} \hat{\alpha}_i = 0,$$
$$[\hat{\alpha}_i, \hat{n}_{ei}] = [\hat{\alpha}_i, \hat{n}_{hi}] = \hat{\alpha}_i, \text{ and} \quad (3)$$
$$[\hat{\alpha}_i^+, \hat{n}_{ei}] = [\hat{\alpha}_i^+, \hat{n}_{hi}] = -\hat{\alpha}_i^+.$$

**II-D. Coupling Strengths of Photons**

In dipole approximation [28], the coupling strength of exciton to photon has the form

$$\gamma_{pi} = i\frac{eE_i}{\hbar}\sqrt{\frac{2\pi\hbar}{\varepsilon_r \omega_p V}} \vec{\varepsilon} \cdot \langle \phi_i(r)|\vec{r}|\phi_i(r)\rangle, \quad (4)$$

where $E_i$ and $\phi_i(r)$ are the exciton energy of the state i and its associated wavefunction (in relative coordinate), respectively, $\vec{\varepsilon}$ is the polarization vector of the photon field, $\omega_p$ is the photon frequency with the mode p, e is the bare electron charge, and $\varepsilon_r$ is the relative dielectric constant of dot's material. The coupling strength Eq. (4) gives the emission rate [28]

$$2\Gamma_i = \frac{4}{3}\frac{e^2 \omega_i^3}{\varepsilon_r \hbar c^3}|\langle \phi_i|\vec{r}|\phi_i\rangle|^2, \quad (5)$$

with $\omega_i = E_i/\hbar$ and the light speed c, $\Gamma_i$ is defined in Sec. III-D. In a disc-like parabolic QD with the identical confinement strengths $\omega_l$ and $\omega_t$ to both electron and hole in the longitudinal and transverse directions, respectively, and assumed that the dot size is much smaller than the Bohr radius of the exciton, i.e. $\omega_l, \omega_t \gg V_{Coul}$ with the Coulomb potential $V_{Coul}$, the exciton's dipole moment (DM) can be written as $|\vec{d}_i| \equiv e|\langle \phi_i(r)|\vec{r}|\phi_i(r)\rangle| = el_c$, $l_c = \sqrt{\hbar/\omega_t \mu^*}$ is the effective cyclotron radius of exciton, and $\mu^*$ is the exciton's reduced mass.

To demonstrate the optical properties of the ground and the first-excited states of a QDM, consider a coupled-QD system with the bare electron (hole) energies $E_{e1(h1)}$ and $E_{e2(h2)}$ to the dots 1 and 2, respectively, and with the tunnel strength $t_{e(h)}$ between the dots. The tunneling t and the Coulomb interaction mix the single-QD states of both electron ($|i\rangle_e$, i denotes the dot 1 or 2) and hole ($|i\rangle_h$). In the regime that the confinement and the tunnel splitting are predominant over the Coulomb effect, a narrow barrier, the energies and their associated wave-functions of the ground and the excited state for the electron (hole) can be approximated as

$$E_{e(h),g} = 1/2[(E_{e(h)1} + E_{e(h)2}) \mp \sqrt{\Delta E_{e(h)}^2 + 4t_{e(h)}^2}], \quad (6)$$

$$|g\rangle_{e(h)} = \frac{1}{\sqrt{1+\eta_{e(h)g}^2}}[\eta_{e(h)}|1\rangle_{e(h)} + |2\rangle_{e(h)}], \quad (7)$$

and

$$E_{e(h),ex} = 1/2[(E_{e(h)1} + E_{e(h)2}) \pm \sqrt{\Delta E_{e(h)}^2 + 4t_{e(h)}^2}], \quad (8)$$

$$|ex\rangle_{e(h)} = \frac{1}{\sqrt{1+\chi^2_{e(h)g}}}\left[\chi_{e(h)}|1\rangle_{e(h)} + |2\rangle_{e(h)}\right] \tag{9}$$

respectively, where $E_{e(h)1} = \hbar\omega_{le(h)}/2 + \hbar\omega_{te(h)}$, $\eta_{e(h)} = \frac{\Delta E_{e(h)} \pm \sqrt{\Delta E^2_{e(h)} + 4t^2_{e(h)}}}{2t_{e(h)}}$,

$\chi_{e(h)} = \frac{\Delta E_{e(h)} \mp \sqrt{\Delta E^2_{e(h)} + 4t^2_{e(h)}}}{2t_{e(h)}}$, $\Delta E_{e(h)} = E_{e(h)2} - E_{e(h)1}$, and $\omega_{le(h)}$ and $\omega_{te(h)}$ are the confinement strengths of electron (hole) in the growth and transverse directions. The QDM states of the exciton may have the forms of $|S1\rangle = |g\rangle_e|g\rangle_h$, $|S2\rangle = |g\rangle_e|ex\rangle_h$, $|S3\rangle = |ex\rangle_e|g\rangle_h$, and $|S4\rangle = |ex\rangle_e|ex\rangle_h$. In double-oscillator approximation and $E_{e(h)1}, E_{e(h)2} \gg \Delta E_{e(h)}$, where $\Delta E_{e(h)}$ is mainly caused by the fluctuation of the dot's high, the tunneling strength of the electron (hole) can be estimated by $t_{e(h)} = \exp(-\alpha^2_{e(h)})\hbar\omega_{le(h)}\alpha_{e(h)}/\sqrt{\pi}$ with the dimensionless distance $\alpha_{e(h)} = a/l_{e(h)}$ between the oscillator centers located at $\pm a$, $l_{e(h)} = \sqrt{\hbar/m^*_{e(hl)}\omega_{le(h)}}$, and $m^*_{e(hl)}$ the effective mass of electron (hole in growth direction) [28].

The DMs of these states including the effect of indirect exciton can be written as

$$|\vec{d}_{S1}| = e\frac{|(1+\eta_e\eta_h)l_c + \sqrt{l^2_c + 4a^2}(\eta_e + \eta_h)C\exp[-2a^2/(l^2_e + l^2_h)]|}{\sqrt{1+\eta^2_e}\sqrt{1+\eta^2_h}}, \tag{10}$$

$$|\vec{d}_{S2}| = e\frac{|(1+\eta_e\chi_h)l_c + \sqrt{l^2_c + 4a^2}(\eta_e + \chi_h)C\exp[-2a^2/(l^2_e + l^2_h)]|}{\sqrt{1+\eta^2_e}\sqrt{1+\chi^2_h}}, \tag{11}$$

$$|\vec{d}_{S3}| = e\frac{|(1+\chi_e\eta_h)l_c + \sqrt{l^2_c + 4a^2}(\chi_e + \eta_h)C\exp[-2a^2/(l^2_e + l^2_h)]|}{\sqrt{1+\chi^2_e}\sqrt{1+\eta^2_h}}, \text{ and} \tag{12}$$

$$|\vec{d}_{S4}| = e\frac{|(1+\chi_e\chi_h)l_c + \sqrt{l^2_c + 4a^2}(\chi_e + \chi_h)C\exp[-2a^2/(l^2_e + l^2_h)]|}{\sqrt{1+\chi^2_e}\sqrt{1+\chi^2_h}}, \tag{13}$$

where the effective confined strength $\omega_t$ of the exciton can be estimated by $\omega_t = \sqrt{(m^*_e\omega^2_{te} + m^*_{ht}\omega^2_{th})/(m^*_e + m^*_{ht})}$, $C = \sqrt{2l_el_h/(l^2_e + l^2_h)}$, and $m^*_{ht}$ is the effective mass of hole in the transverse direction. The first term in Eqs. (10)-(13) is resulted from the effect of intra-dot (direct) exciton while the last one from the inter-dot (indirect) exciton. The confinement strengths of electron (hole) in growth and transverse directions can be approximately estimated by $\omega_{le(h)} = \sqrt{8V_{co(bo)}/m^*_{e(hl)}h^2}$ and $\omega_{te(h)} = \sqrt{8V_{co(bo)}/m^*_{e(ht)}R^2}$, respectively, where $V_{co(bo)}$ is the conduction (valence) band-offset, h is dot's high, and R is dot's radius.

For $m^*_e = 0.081m_0$, $m^*_{hl} = 0.34m_0$, $m^*_{ht} = 0.153m_0$, $V_{co} = 0.68eV$, $V_{bo} = 0.1eV$, $E_{g(dot)} = 0.73eV$, $h = 3.5nm$, $R = 17nm$, and $\Delta E_h = -\Delta E_e/5$, the calculated results of the energy levels of exciton and their DMs with respect to $\Delta E_e$ and $2a$ (the distance between the dots) are plotted in Figs. 2 and 3, respectively. For a symmetric structure, $\eta_i=1$ and $\chi_i=-1$, the DMs of the states |S2> and |S3> become zero, a dark-exciton state, as shown in Fig. 2(a). With increasing its asymmetry, $\Delta E_e$ increases, the distribution of the ground state of electron and hole tends towards the low energy side, and that of the excited state towards the other side. These inhomogeneous distributions of the electron and hole result to increase the DMs of |S2> and |S3>. For a long inter-dot distance, the coupling between the dots reduces exponentially, Fig. 3(c), and the energy differences between |S1> and |S2> and between |S3> and |S4> are reasonable to reduce to $\Delta E_h$, Fig. 3(b). In this situation, the exciton states |S1> and |S4> approach the isolated QD's states and their DMs reduce to the QD's ones, while the states |S2> and |S3> become a pure indirect exciton and their DMs approach zero, Fig. 3(a), because the inter-dot wavefunction overlap of the electron and hole approaches zero. At short distance, the tunnel splitting that is greater than the energy difference caused by the asymmetry of the dots, Fig. 3(c), dominates the DMs of the states |S2> and |S3>. In this situation, the property of the system tends towards the symmetry one, and the DMs of the states |S2> and |S3> shrink, Fig. 3(a). The increase of the tunneling strengths increases the gaps between these four states, Fig. 3(b). Figure (4) shows the transition rates of these four states with respect to the inter-dot distance for $\Delta E_e = 15meV$ and $\Delta E_{eh} = 3meV$. It clearly displays that the transition rates of both states |S2> and |S3> are one order smaller than the states |S1> and |S4> at whole distance, i.e., these states have the lifetime much longer than the states of |S1> and |S4>. It is worthy to be noted that the transition rate of the state |S1> increases nearly two times that of the single QD (~0.61μeV), which takes the value ~1.16μeV, as deceased the inter-dot distance to ~2.5nm due to the contribution of the indirect exciton. The transition rate of QDM (QD) gives the exciton decay-time of ~0.57ns (~1.1ns). The calculated results show that the model used here is qualitatively correct.

**II-E. Coupling Strengths of LO phonons**

In the estimation of LO-phonon coupling strength, consider the LO-phonons confined in a cubic quantum box with size $\zeta$ and a pressure-free boundary condition. The coupling strength $\gamma_q$ of mode q has the form proposed in Ref[29]

$$\gamma_q = \frac{4e}{|\vec{q}|}\sqrt{\frac{\pi\hbar\omega_{LO}}{\zeta^3}}\left(\frac{1}{\varepsilon_\infty}-\frac{1}{\varepsilon_0}\right)\sin(q_x x)\sin(q_y y)\sin(q_z z), \qquad (14)$$

where $\varepsilon_\infty$ and $\varepsilon_0$ are the high frequency and static dielectric constants of dot's material. The total coupling strength has been estimated for GaAs QD [29] to be

$$\gamma_0 = \sqrt{\sum_q \gamma_q^2} = \frac{0.35\hbar\omega_{LO}}{\sqrt{\zeta}}. \qquad (15)$$

For InAs QD, the factor 0.35 is simply rescaled by $(\varepsilon_\infty^{-1}-\varepsilon_0^{-1})_{InAs}/(\varepsilon_\infty^{-1}-\varepsilon_0^{-1})_{GaAs}$ and takes the value ~0.374. For $\zeta \cong 8.56$nm that corresponds to the volume of a QD with the diameter 20nm and high 2nm, and $\hbar\omega_{LO} = 36meV$, the total strength is estimated to be ~4.6meV. This result will be used in later discussions.

## III. Equations of Motion

**III-A. Approximation Scheme**

The kinetics of the system can be described by a set of master equations obtained by applying the EOM approach, i.e.

$$i\partial_t <\hat{O}> = <[\hat{O},\hat{H}]>, \qquad (16)$$

to the density operators of exciton, photon, and phonon (in units of $\hbar = 1$) and taking the thermal average over the operator and the commutator. In the derivation of these equations, the following approximations are made:

(1) The small interacting strength between excitons and photons allows us to approximately correct the optical transitions to the second order, i.e., to the orders of $|\gamma_{pg}|^2$ and $|\gamma_{pe}|^2$. In such an approximation, the terms with non-conserved photon momentum (such as $<\cdots \hat{a}_p^+ \hat{a}_{p'}>$) and with multiple-photon processes, such as $<\cdots \hat{a}_p^+ \hat{a}_{p'}^+>$, are ignored.

(2) The terms with the multiple poles, such as $1/[(s-\omega_a)+i\delta][(s-\omega_b)+i\delta]$ in the

s-plane (a dual space in the Laplace transformation), that are well separated, i.e., $|\omega_a - \omega_b| \gg \delta$ with δ being the broadening width, are neglected.

(3) In the case of low-power excitation, where only one exciton presents in the system, the terms with multiple-exciton processes, for examples $<\cdots \hat{\alpha}_g \hat{\alpha}_e>$, $<\cdots \hat{\alpha}_g^+ \hat{n}_{ee}>$, and $<\cdots \hat{\alpha}_g^+ \hat{\alpha}_g \hat{\alpha}_e>$, are truncated.

Under these approximations, an analytical solution to the equation set is possible.

### III-B. EOMs of Density Matrix

To describe the time evolution of the densities with multiple varieties, the equation (16) is first applied to the lowest-order diagonal elements of the densities and yields the equations

$$i\partial_t <\hat{a}_p^+ \hat{a}_p> = \gamma_{pi} <\hat{a}_p^+ \hat{\alpha}_i> - \gamma_{pi}^* <\hat{\alpha}_i^+ \hat{a}_p>, \tag{17}$$

$$i\partial_t <\hat{\alpha}_g^+ \hat{\alpha}_g> = -\gamma_{pg} <\hat{a}_p^+ \hat{\alpha}_g> + \gamma_{pg}^* <\hat{\alpha}_g^+ \hat{a}_p> + \gamma_1 <\hat{d}_1^+ \hat{\alpha}_g^+ \hat{\alpha}_e> - \gamma_1^* <\hat{d}_1 \hat{\alpha}_e^+ \hat{\alpha}_g>, \tag{18}$$

$$i\partial_t <\hat{\alpha}_e^+ \hat{\alpha}_e> = -\gamma_{pe} <\hat{a}_p^+ \hat{\alpha}_e> + \gamma_{pe}^* <\hat{\alpha}_e^+ \hat{a}_p> - \gamma_1 <\hat{d}_1^+ \hat{\alpha}_g^+ \hat{\alpha}_e> + \gamma_1^* <\hat{d}_1 \hat{\alpha}_e^+ \hat{\alpha}_g>. \tag{19}$$

The repeated subscripts that do not appear in the left-hand side of the EOMs imply a summation to be performed over the notation, for examples $\gamma_{pg} <\hat{a}_p^+ \hat{\alpha}_i> = \sum_{p,i} \gamma_{pi} <\hat{a}_p^+ \hat{\alpha}_i>$ and $\gamma_1 <\hat{d}_1^+ \hat{\alpha}_g^+ \hat{\alpha}_e> = \sum_{q_1} \gamma_{q_1} <\hat{d}_{q_1}^+ \hat{\alpha}_g^+ \hat{\alpha}_e>$. The time evolution of the diagonal densities involving the mechanisms of particle transition generates the off-diagonal densities shown in Eqs. (17)-(19). To know the dynamics of the off-diagonal densities, one would apply Eq. (16) again to these new generated terms and gives

$$i\partial_t <\hat{a}_p^+ \hat{\alpha}_g> \cong (E_g - \omega_p) <\hat{a}_p^+ \hat{\alpha}_g> + \gamma_{pg}^* <\hat{a}_p^+ \hat{a}_p> \\ - \gamma_{pg}^* <\hat{\alpha}_g^+ \hat{\alpha}_g> + \gamma_1 <\hat{d}_1^+ \hat{a}_p^+ \hat{\alpha}_e>, \tag{20}$$

$$i\partial_t <\hat{a}_p^+ \hat{\alpha}_e> \cong (E_e - \omega_p) <\hat{a}_p^+ \hat{\alpha}_e> + \gamma_{pe}^* <\hat{a}_p^+ \hat{a}_p> \\ - \gamma_{pe}^* <\hat{\alpha}_e^+ \hat{\alpha}_e> + \gamma_1^* <\hat{d}_1 \hat{a}_p^+ \hat{\alpha}_g>, \tag{21}$$

$$i\partial_t <\hat{d}_1^+\hat{\alpha}_g^+\hat{\alpha}_e> \cong (E_e - E_g - \omega_{LO})<\hat{d}_1^+\hat{\alpha}_g^+\hat{\alpha}_e> + \gamma_2^* <\hat{d}_1^+\hat{d}_2\hat{\alpha}_g^+\hat{\alpha}_g>$$
$$-\gamma_2^* <\hat{d}_2\hat{d}_1^+\hat{\alpha}_e^+\hat{\alpha}_e> - \gamma_{pg} <\hat{d}_1^+\hat{a}_p^+\hat{\alpha}_e> + \gamma_{pe}^* <\hat{d}_1^+\hat{\alpha}_g^+\hat{a}_p>, \quad (22)$$

$$i\partial_t <\hat{d}_1^+\hat{a}_p^+\hat{\alpha}_e> \cong (E_e - \omega_p - \omega_{LO})<\hat{d}_1^+\hat{a}_p^+\hat{\alpha}_e> + \gamma_{pe}^* <\hat{d}_1^+\hat{a}_p^+\hat{a}_p>$$
$$-\gamma_{pe}^* <\hat{d}_1^+\hat{\alpha}_e^+\hat{\alpha}_e> + \gamma_2^* <\hat{d}_1^+\hat{d}_2\hat{a}_p^+\hat{\alpha}_g> - \gamma_{pg}^* <\hat{d}_1^+\hat{\alpha}_g^+\hat{\alpha}_e>, \quad (23)$$
$$\cong (E_e - \omega_p - \omega_{LO})<\hat{d}_1^+\hat{a}_p^+\hat{\alpha}_e> + \gamma_2^* <\hat{d}_1^+\hat{d}_2\hat{a}_p^+\hat{\alpha}_g> - \gamma_{pg}^* <\hat{d}_1^+\hat{\alpha}_g^+\hat{\alpha}_e>$$

where $\omega_p$ ($\omega_{LO}$) is the photon (LO-phonon) energy, $E_i$ is the energy of the bare exciton in the state i. In the derivation, the approximation rules stated in the last section and the identities of Eq. (3) are applied. Note that the EOMs for the off-diagonal densities have a nonzero pole revealed in the first term of these equations. It can be easily checked that the terms of $<d_1^+ a_p^+ a_p>$ and $<d_1^+ \alpha_g^+ \alpha_g>$ in Eq. (23) have the pole at $\omega_{LO}$ that is much different from the pole $E_e - \omega_p - \omega_{LO}$ for $<d_1^+ a_p^+ \alpha_e>$, and thus can be further ignored due to the approximation rule (2).

At the first glance, the EQMs seem likely to be endless because the new higher-order terms are generated when one applies Eq. (16) to the new terms over and over again. Fortunately, there exists recurrence relations to the endless EOMs under the approximations used in this work, which are

$$i\partial_t <(\hat{d}_\bullet \hat{d}_\bullet^+)^j \hat{a}_p^+ \hat{a}_p> = \gamma_{pi} <(\hat{d}_\bullet \hat{d}_\bullet^+)^j \hat{a}_p^+ \hat{\alpha}_i> - \gamma_{pi}^* <(\hat{d}_\bullet \hat{d}_\bullet^+)^j \hat{\alpha}_i^+ \hat{a}_p>, \quad (24)$$

$$i\partial_t <(\hat{d}_\bullet^+ \hat{d}_\bullet)^j \hat{a}_p^+ \hat{a}_p> = \gamma_{pi} <(\hat{d}_\bullet^+ \hat{d}_\bullet)^j \hat{a}_p^+ \hat{\alpha}_i> - \gamma_{pi}^* <(\hat{d}_\bullet^+ \hat{d}_\bullet)^j \hat{\alpha}_i^+ \hat{a}_p>, \quad (25)$$

$$i\partial_t <(\hat{d}_\bullet^+ \hat{d}_\bullet)^j \hat{\alpha}_g^+ \hat{\alpha}_g> = -\gamma_{pg} <(\hat{d}_\bullet^+ \hat{d}_\bullet)^j \hat{a}_p^+ \hat{\alpha}_g> + \gamma_{pg}^* <(\hat{d}_\bullet^+ \hat{d}_\bullet)^j \hat{\alpha}_g^+ \hat{a}_p>$$
$$+\gamma_{2j+1} <(\hat{d}_\bullet^+ \hat{d}_\bullet)^j \hat{d}_{2j+1}^+ \hat{\alpha}_g^+ \hat{\alpha}_e> - \gamma_{2j+1}^* <\hat{d}_{2j+1}(\hat{d}_\bullet^+ \hat{d}_\bullet)^j \hat{\alpha}_e^+ \hat{\alpha}_g>, \quad (26)$$

$$i\partial_t <(\hat{d}_\bullet \hat{d}_\bullet^+)^j \hat{\alpha}_e^+ \hat{\alpha}_e> = -\gamma_{pe} <(\hat{d}_\bullet \hat{d}_\bullet^+)^j \hat{a}_p^+ \hat{\alpha}_e> + \gamma_{pe}^* <(\hat{d}_\bullet \hat{d}_\bullet^+)^j \hat{\alpha}_e^+ \hat{a}_p>$$
$$-\gamma_{2j+1} <\hat{d}_{2j+1}^+ (\hat{d}_\bullet \hat{d}_\bullet^+)^j \hat{\alpha}_g^+ \hat{\alpha}_e> + \gamma_{2j+1}^* <(\hat{d}_\bullet \hat{d}_\bullet^+)^j \hat{d}_{2j+1} \hat{\alpha}_e^+ \hat{\alpha}_g>, \quad (27)$$

$$i\partial_t <(\hat{d}_\bullet^+ \hat{d}_\bullet)^j \hat{d}_{2j+1}^+ \hat{\alpha}_g^+ \hat{\alpha}_e> = (E_e - E_g - \omega_{LO})<(\hat{d}_\bullet^+ \hat{d}_\bullet)^j \hat{d}_{2j+1}^+ \hat{\alpha}_g^+ \hat{\alpha}_e>$$
$$-\gamma_{pg} <(\hat{d}_\bullet^+ \hat{d}_\bullet)^j \hat{d}_{2j+1}^+ \hat{a}_p^+ \hat{\alpha}_e> + \gamma_{pe}^* <(\hat{d}_\bullet^+ \hat{d}_\bullet)^j \hat{d}_{2j+1}^+ \hat{\alpha}_g^+ \hat{a}_p>, \quad (28)$$
$$+\gamma_{2j+2}^* <(\hat{d}_\bullet^+ \hat{d}_\bullet)^{j+1} \hat{\alpha}_g^+ \hat{\alpha}_g> - \gamma_{2j+2}^* <(\hat{d}_\bullet \hat{d}_\bullet^+)^{j+1} \hat{\alpha}_e^+ \hat{\alpha}_e>$$

$$i\partial_t <(\hat{d}_\bullet^+\hat{d}_\bullet)^j \hat{a}_p^+\hat{\alpha}_g> = (E_g - \omega_p)<(\hat{d}_\bullet^+\hat{d}_\bullet)^j \hat{a}_p^+\hat{\alpha}_g> + \gamma_{pg}^* <(\hat{d}_\bullet^+\hat{d}_\bullet)^j \hat{a}_p^+\hat{a}_p>$$
$$-\gamma_{pg}^* <(\hat{d}_\bullet^+\hat{d}_\bullet)^j \hat{\alpha}_g^+\hat{\alpha}_g> + \gamma_{2j+1} <(\hat{d}_\bullet^+\hat{d}_\bullet)^j \hat{d}_{2j+1}^+\hat{a}_p^+\hat{\alpha}_e>,$$ (29)

$$i\partial_t <(\hat{d}_\bullet\hat{d}_\bullet^+)^j \hat{a}_p^+\hat{\alpha}_e> = (E_e - \omega_p)<(\hat{d}_\bullet\hat{d}_\bullet^+)^j \hat{a}_p^+\hat{\alpha}_e> + \gamma_{pe}^* <(\hat{d}_\bullet\hat{d}_\bullet^+)^j \hat{a}_p^+\hat{a}_p>$$
$$-\gamma_{pe}^* <(\hat{d}_\bullet\hat{d}_\bullet^+)^j \hat{\alpha}_e^+\hat{\alpha}_e> + \gamma_{2j+1}^* <(\hat{d}_\bullet\hat{d}_\bullet^+)^j \hat{d}_{2j+1}\hat{a}_p^+\hat{\alpha}_g>,$$ (30)

$$i\partial_t <(\hat{d}_\bullet^+\hat{d}_\bullet)^j \hat{d}_{2j+1}^+\hat{a}_p^+\hat{\alpha}_e> = (E_e - \omega_p - \omega_{LO})<(\hat{d}_\bullet^+\hat{d}_\bullet)^j \hat{d}_{2j+1}^+\hat{a}_p^+\hat{\alpha}_e>$$
$$+\gamma_{2j+2}^* <(\hat{d}_\bullet^+\hat{d}_\bullet)^{j+1}\hat{a}_p^+\hat{\alpha}_g> - \gamma_{pg}^* <(\hat{d}_\bullet^+\hat{d}_\bullet)^j \hat{d}_{2j+1}^+\hat{\alpha}_g^+\hat{\alpha}_e>,$$ (31)

$$i\partial_t <(\hat{d}_\bullet^+\hat{d}_\bullet)^j \hat{d}_{2j+1}^+\hat{\alpha}_g^+\hat{a}_p> = (\omega_p - E_g - \omega_{LO})<(\hat{d}_\bullet^+\hat{d}_\bullet)^j \hat{d}_{2j+1}^+\hat{\alpha}_g^+\hat{a}_p>$$
$$-\gamma_{2j+2}^* <(\hat{d}_\bullet\hat{d}_\bullet^+)^{j+1}\hat{\alpha}_e^+\hat{a}_p> + \gamma_{pe} <(\hat{d}_\bullet^+\hat{d}_\bullet)^j \hat{d}_{2j+1}^+\hat{\alpha}_g^+\hat{\alpha}_e>,$$ (32)

with $(\hat{d}_\bullet^+\hat{d}_\bullet)^j = \hat{d}_1^+\hat{d}_2\cdots\hat{d}_{2j-1}^+\hat{d}_{2j}$. For j=0, the equations (24)-(27) become Eqs. (17)-(19). Solving these equations with a proper initial condition yields the time evolution to the population densities.

### III-C. Initial Conditions

The solutions to Eqs. (24)-(32) are governed by the initial conditions of the densities. In the sudden approximation, the Hamiltonian changes abruptly due to the sudden change of the populations at t=0, and, hence, one would expect that the wave-function does not change much at the initial time and the system is approximately retained in its equilibrium. In this situation, the thermal average of the off-diagonal densities approaches zero [30], while the diagonal densities have the form, for example,

$$<(\hat{d}_\bullet^+\hat{d}_\bullet)^j \hat{\alpha}_i^+\hat{\alpha}_i>_0 = \delta_{\bullet,\bullet}<(\hat{d}_\bullet^+\hat{d}_\bullet)^j>_0 N_i(0).$$ (33)

The delta function with the subscript • means that the phonon operators have to be paired in all possible ways, such as the term with four phonon operators [30]

$$\delta_{\bullet,\bullet}<\hat{d}_1^+\hat{d}_2\hat{d}_3^+\hat{d}_4>_0 = (\delta_{1,2}\delta_{3,4} + \delta_{1,4}\delta_{2,3})<\hat{d}_1^+\hat{d}_2\hat{d}_3^+\hat{d}_4>_0 = N_B^2 + N_B(1+N_B)$$ (34)

with the Planck's distribution $N_B$ and the initial number of photons (excitons)

$N_p(0) \equiv <\hat{a}_p^+ \hat{a}_p>_0$ ($N_i(0) \equiv <\hat{\alpha}_i^+ \hat{\alpha}_i>_0$). The first term in Eq. (34) is resulted from the direct LO-phonon process, while the second one from the exchange LO-phonon.

**III-D. EOMs in S-plane**

In solving the differential equations (24)-(32), it is advantage to transfer these equations into algebra problem by using the Laplace transformation, defined as $\overline{N}(s) = \int_0^\infty N(t)\exp(-st)dt$, with the initial conditions stated in the last section. Laplace transforms of the equations yield

$$is\overline{<(\hat{d}_\bullet\hat{d}_\bullet^+)^j \hat{a}_p^+ \hat{a}_p>} = i\delta_{\bullet,\bullet}<(\hat{d}_\bullet\hat{d}_\bullet^+)^j \hat{a}_p^+ \hat{a}_p>_0 \\ +\gamma_{pe}\overline{<(\hat{d}_\bullet\hat{d}_\bullet^+)^j \hat{a}_p^+ \hat{\alpha}_e>} -\gamma_{pe}^*\overline{<(\hat{d}_\bullet\hat{d}_\bullet^+)^j \hat{\alpha}_e^+ \hat{a}_p>}, \quad (24a)$$

$$is\overline{<(\hat{d}_\bullet^+\hat{d}_\bullet)^j \hat{a}_p^+ \hat{a}_p>} = i\delta_{\bullet,\bullet}<(\hat{d}_\bullet^+\hat{d}_\bullet)^j \hat{a}_p^+ \hat{a}_p>_0 \\ +\gamma_{pg}\overline{<(\hat{d}_\bullet^+\hat{d}_\bullet)^j \hat{a}_p^+ \hat{\alpha}_g>} -\gamma_{pg}^*\overline{<(\hat{d}_\bullet^+\hat{d}_\bullet)^j \hat{\alpha}_g^+ \hat{a}_p>}, \quad (25a)$$

$$is\overline{<(\hat{d}_\bullet^+\hat{d}_\bullet)^j \hat{\alpha}_g^+ \hat{\alpha}_g>} = i\delta_{\bullet,\bullet}<(\hat{d}_\bullet^+\hat{d}_\bullet)^j \hat{\alpha}_g^+ \hat{\alpha}_g>_0 \\ -\gamma_{pg}\overline{<(\hat{d}_\bullet^+\hat{d}_\bullet)^j \hat{a}_p^+ \hat{\alpha}_g>} +\gamma_{pg}^*\overline{<(\hat{d}_\bullet^+\hat{d}_\bullet)^j \hat{\alpha}_g^+ \hat{a}_p>} \\ +\gamma_{2j+1}\overline{<(\hat{d}_\bullet^+\hat{d}_\bullet)^j \hat{d}_{2j+1}^+ \hat{\alpha}_g^+ \hat{\alpha}_e>} -\gamma_{2j+1}^*\overline{<\hat{d}_{2j+1}(\hat{d}_\bullet^+\hat{d}_\bullet)^j \hat{\alpha}_e^+ \hat{\alpha}_g>}, \quad (26a)$$

$$is\overline{<(\hat{d}_\bullet\hat{d}_\bullet^+)^j \hat{\alpha}_e^+ \hat{\alpha}_e>} = i\delta_{\bullet,\bullet}<(\hat{d}_\bullet\hat{d}_\bullet^+)^j \hat{\alpha}_e^+ \hat{\alpha}_e>_0 \\ -\gamma_{pe}\overline{<(\hat{d}_\bullet\hat{d}_\bullet^+)^j \hat{a}_p^+ \hat{\alpha}_e>} +\gamma_{pe}^*\overline{<(\hat{d}_\bullet\hat{d}_\bullet^+)^j \hat{\alpha}_e^+ \hat{a}_p>} \\ -\gamma_{2j+1}\overline{<\hat{d}_{2j+1}^+(\hat{d}_\bullet\hat{d}_\bullet^+)^j \hat{\alpha}_g^+ \hat{\alpha}_e>} +\gamma_{2j+1}^*\overline{<(\hat{d}_\bullet\hat{d}_\bullet^+)^j \hat{d}_{2j+1} \hat{\alpha}_e^+ \hat{\alpha}_g>}, \quad (27a)$$

$$is\overline{<(\hat{d}_\bullet^+\hat{d}_\bullet)^j \hat{d}_{2j+1}^+ \hat{\alpha}_g^+ \hat{\alpha}_e>} = (E_e - E_g - \omega_{LO})\overline{<(\hat{d}_\bullet^+\hat{d}_\bullet)^j \hat{d}_{2j+1}^+ \hat{\alpha}_g^+ \hat{\alpha}_e>} \\ -\gamma_{pg}\overline{<(\hat{d}_\bullet^+\hat{d}_\bullet)^j \hat{d}_{2j+1}^+ \hat{a}_p^+ \hat{\alpha}_e>} +\gamma_{pe}^*\overline{<(\hat{d}_\bullet^+\hat{d}_\bullet)^j \hat{d}_{2j+1}^+ \hat{\alpha}_g^+ \hat{a}_p>}, \quad (28a) \\ +\gamma_{2j+2}^*\overline{<(\hat{d}_\bullet^+\hat{d}_\bullet)^{j+1} \hat{\alpha}_g^+ \hat{\alpha}_g>} -\gamma_{2j+2}^*\overline{<(\hat{d}_\bullet\hat{d}_\bullet^+)^{j+1} \hat{\alpha}_e^+ \hat{\alpha}_e>}$$

$$is\overline{<(\hat{d}_\bullet^+\hat{d}_\bullet)^j \hat{a}_p^+ \hat{\alpha}_g>} = (E_g - \omega_p)\overline{<(\hat{d}_\bullet^+\hat{d}_\bullet)^j \hat{a}_p^+ \hat{\alpha}_g>} +\gamma_{pg}^*\overline{<(\hat{d}_\bullet^+\hat{d}_\bullet)^j \hat{a}_p^+ \hat{a}_p>} \\ -\gamma_{pg}^*\overline{<(\hat{d}_\bullet^+\hat{d}_\bullet)^j \hat{\alpha}_g^+ \hat{\alpha}_g>} +\gamma_{2j+1}\overline{<(\hat{d}_\bullet^+\hat{d}_\bullet)^j \hat{d}_{2j+1}^+ \hat{a}_p^+ \hat{\alpha}_e>}, \quad (29a)$$

$$is\overline{<(\hat{d}_\bullet\hat{d}_\bullet^+)^j\hat{a}_p^+\hat{\alpha}_e>} = (E_e - \omega_p)\overline{<(\hat{d}_\bullet\hat{d}_\bullet^+)^j\hat{a}_p^+\hat{\alpha}_e>} + \gamma_{pe}^*\overline{<(\hat{d}_\bullet\hat{d}_\bullet^+)^j\hat{a}_p^+\hat{a}_p>}$$
$$-\gamma_{pe}^*\overline{<(\hat{d}_\bullet\hat{d}_\bullet^+)^j\hat{\alpha}_e^+\hat{\alpha}_e>} + \gamma_{2j+1}^*\overline{<(\hat{d}_\bullet\hat{d}_\bullet^+)^j\hat{d}_{2j+1}\hat{a}_p^+\hat{\alpha}_g>},\quad(30a)$$

$$is\overline{<(\hat{d}_\bullet^+\hat{d}_\bullet)^j\hat{d}_{2j+1}^+\hat{a}_p^+\hat{\alpha}_e>} = (E_e - \omega_p - \omega_{LO})\overline{<(\hat{d}_\bullet^+\hat{d}_\bullet)^j\hat{d}_{2j+1}^+\hat{a}_p^+\hat{\alpha}_e>}$$
$$+\gamma_{2j+2}^*\overline{<(\hat{d}_\bullet^+\hat{d}_\bullet)^{j+1}\hat{a}_p^+\hat{\alpha}_g>} - \gamma_{pg}^*\overline{<(\hat{d}_\bullet^+\hat{d}_\bullet)^j\hat{d}_{2j+1}^+\hat{\alpha}_g^+\hat{\alpha}_e>},\quad(31a)$$

$$is\overline{<(\hat{d}_\bullet^+\hat{d}_\bullet)^j\hat{d}_{2j+1}^+\hat{\alpha}_g^+\hat{a}_p>} = (\omega_p - E_g - \omega_{LO})\overline{<(\hat{d}_\bullet^+\hat{d}_\bullet)^j\hat{d}_{2j+1}^+\hat{\alpha}_g^+\hat{a}_p>}$$
$$-\gamma_{2j+2}^*\overline{<(\hat{d}_\bullet\hat{d}_\bullet^+)^{j+1}\hat{\alpha}_e^+\hat{a}_p>} + \gamma_{pe}\overline{<(\hat{d}_\bullet^+\hat{d}_\bullet)^j\hat{d}_{2j+1}^+\hat{\alpha}_g^+\hat{\alpha}_e>}.\quad(32a)$$

Algebraically solving these equations, one obtains the solutions to the density matrix in the s-domain.

In the indirect excitation, the initial number of photons with the energy around the ground and the excited states of the exciton is zero, i.e. $N_p(0)=0$. This condition yields the solutions to Eqs. (26a) and (27a) with the recurrence relations between the orders j and j+1

$$\overline{<(\hat{d}_\bullet^+\hat{d}_\bullet)^j\hat{\alpha}_g^+\hat{\alpha}_g>} = \frac{\delta_{\bullet,\bullet}<(\hat{d}_\bullet^+\hat{d}_\bullet)^j>_0 N_g(0)}{(s+2\Gamma_g)}$$
$$-\frac{2\gamma_{2j+1}^*\gamma_{2j+2}(s+\Gamma_g+\Gamma_e)[\overline{<(\hat{d}_\bullet^+\hat{d}_\bullet)^{j+1}\hat{\alpha}_g^+\hat{\alpha}_g>} - \overline{<(\hat{d}_\bullet\hat{d}_\bullet^+)^{j+1}\hat{\alpha}_e^+\hat{\alpha}_e>}]}{(s+2\Gamma_g)[(s+\Gamma_g+\Gamma_e)^2+(E_e-E_g-\omega_{LO}-E_{Ng}+E_{Ne})^2]},\quad(35)$$

$$\overline{<(\hat{d}_\bullet\hat{d}_\bullet^+)^j\hat{\alpha}_e^+\hat{\alpha}_e>} = \frac{\delta_{\bullet,\bullet}<(\hat{d}_\bullet\hat{d}_\bullet^+)^j>_0 N_e(0)}{(s+2\Gamma_e)}$$
$$+\frac{2\gamma_{2j+1}^*\gamma_{2j+2}(s+\Gamma_g+\Gamma_e)[\overline{<(\hat{d}_\bullet^+\hat{d}_\bullet)^{j+1}\hat{\alpha}_g^+\hat{\alpha}_g>} - \overline{<(\hat{d}_\bullet\hat{d}_\bullet^+)^{j+1}\hat{\alpha}_e^+\hat{\alpha}_e>}]}{(s+2\Gamma_e)[(s+\Gamma_g+\Gamma_e)^2+(E_e-E_g-\omega_{LO}-E_{Ng}+E_{Ne})^2]},\quad(36)$$

where the optical-transition rate is defined as $\Gamma_{g(e)} \equiv \text{Re}\{\sum_p \gamma_{pg(e)}^2/[s-i(E_{g(e)}-\omega_p)]\}$,

and the renormalization energy caused by the excitonic polaron coupling to the photons is given by $E_{Ng(e)} = \text{Im}\{\sum_p \gamma_{pg(e)}^2/[s-i(E_{e(g)}-\omega_p-\omega_{LO})]\}$.

### III-E. Analytical Solutions

Equations (35) and (36) can be solved step by step starting from the lowest-order equations, i.e., j=0. The equations for each order are terminated at the first (initial) term governed by the initial conditions for that order, otherwise, will be related its higher-order ones. There are many combinations in the initial term for j>1 as described in Sec. III-C. The choice of the combinations depends on the last initial term. For example, the initial term exists only for $\delta_{1,2}$ ($\delta_{1,2} <\hat{d}_1^+\hat{d}_2>$) to the first order, there are two combinations to the second order, which are $\delta_{1,2}\delta_{3,4}$ and $\delta_{1,4}\delta_{2,3}$, and so forth to the higher orders. If one counts the contribution from the initial term of j=1, the exchange term ($\delta_{1,4}\delta_{2,3}$) for j=2 disappears. In analogy, one can deduce that only one combination exists, that is

$$\delta_{\bullet,\bullet}<(\hat{d}_\bullet^+\hat{d}_\bullet)^j>_0 = \delta_{1,2}\delta_{3,4}\cdots\delta_{2j-1,2j}<\hat{d}_1^+\hat{d}_2\hat{d}_3^+\hat{d}_4\cdots\hat{d}_{2j-1}^+\hat{d}_{2j}>_0 = N_B^j,$$

for counting the contribution from every order smaller than j. Since the exciton interacts with the LO-phonon in the manner that the exciton emits and then absorbs the same phonon coherently, one may call it as *resonant phonon*.

Iterative substitution of the equations (35) and (36) for the order j+1 into those of the order j yields

$$\begin{aligned}\overline{N}_g(s) &= \overline{<\hat{\alpha}_g^+\hat{\alpha}_g>} \\ &= P_g N_g(0) - 2P_g^2 Q\delta_{1,2}\gamma_1^*\gamma_2[<\hat{d}_1^+\hat{d}_2>_0 N_g(0) - <\hat{d}_1\hat{d}_2^+>_0 N_e(0)] \\ &+ 4P_g^3 Q^2 \delta_{1,2}\delta_{3,4}\gamma_1^*\gamma_2\gamma_3^*\gamma_4[<\hat{d}_1^+\hat{d}_2\hat{d}_3^+\hat{d}_4>_0 N_g(0) - <\hat{d}_1\hat{d}_2^+\hat{d}_3\hat{d}_4^+>_0 N_e(0)] \\ &-+\cdots\end{aligned} \quad (37)$$

and

$$\begin{aligned}\overline{N}_e(s) &= \overline{<\hat{\alpha}_e^+\hat{\alpha}_e>} \\ &= P_e N_e(0) + 2P_e^2 Q\delta_{1,2}\gamma_1^*\gamma_2[<\hat{d}_1^+\hat{d}_2>_0 N_g(0) - <\hat{d}_1\hat{d}_2^+>_0 N_e(0)] \\ &+ 4P_e^3 Q^2 \delta_{1,2}\delta_{3,4}\gamma_1^*\gamma_2\gamma_3^*\gamma_4[<\hat{d}_1^+\hat{d}_2\hat{d}_3^+\hat{d}_4>_0 N_g(0) - <\hat{d}_1\hat{d}_2^+\hat{d}_3\hat{d}_4^+>_0 N_e(0)] \\ &-+\cdots\end{aligned} \quad , (38)$$

where $P_i \equiv 1/(s+2\Gamma_i)$ and $Q \equiv (s+\Gamma_g+\Gamma_e)/[(s+\Gamma_g+\Gamma_e)^2+\Delta^2]$ with the detuning energy $\Delta \equiv E_e - E_g - \omega_{LO} - E_{Ng} + E_{Ne}$.

Applying the initial conditions mentioned above and the identity $1/(1+x) = 1 - x + x^2 - +\cdots$ to the equations (37) and (38) yields $\overline{N}_g(s)$ and $\overline{N}_e(s)$

in the Dyson's form

$$\bar{N}_g(s) = N_g(0)\frac{P_g(1+\gamma_0^2 P_e Q N_B)}{1+(P_g+P_e)\gamma_0^2 Q N_B} + N_e(0)\frac{P_g P_e \gamma_0^2 Q(1+N_B)}{1+(P_g+P_e)\gamma_0^2 Q(1+N_B)} \quad \text{and} \tag{39}$$

$$\bar{N}_e(s) = N_e(0)\frac{P_e[1+\gamma_0^2 P_g Q(1+N_B)]}{1+(P_g+P_e)\gamma_0^2 Q(1+N_B)} + N_g(0)\frac{P_g P_e \gamma_0^2 Q N_B}{1+(P_g+P_e)\gamma_0^2 Q N_B}, \tag{40}$$

with the total strength $\gamma_0 = (\sum_q |\gamma_q|^2)^{1/2}$ of the exciton interacting to all LO-phonon modes. Usually, the term $s+\Gamma_e+\Gamma_g$ in the denominator of Q is negligible, keeping this term results to a RO, because its value (in the order of μeV) is usually much smaller than Δ (in the order of meV) for InGaAs/GaAs QDs and QDMs. Such an approximation that is equivalent to removing the RO by filtered off the high-frequency signal in the responses of time-resolved PL (TRPL), yields $\gamma_0^2 Q \cong 2(s+\Gamma_e+\Gamma_g)\beta^2$ with the dimensionless strength $\beta \equiv \gamma_0/\Delta$. Notably, the square of β times the phonon distribution $N_B$ is the number of LO phonons that clothe an exciton [31].

For T=0, Eqs. (39) and (40) are reduced to

$$\bar{N}_g(s) = N_g(0)P_g + N_e(0)\frac{P_g P_e \gamma_0^2 Q}{1+(P_g+P_e)\gamma_0^2 Q} \quad \text{and} \tag{41}$$

$$\bar{N}_e(s) = N_e(0)\frac{P_e[1+\gamma_0^2 P_g Q]}{1+(P_g+P_e)\gamma_0^2 Q}, \tag{42}$$

where the spontaneous emission gives a nonzero dependence on $N_e(0)$ for both $\bar{N}_g(s)$ and $\bar{N}_e(s)$. For $N_g(0)=1$ and $N_e(0)=0$, the decay rate is $2\Gamma_g$ and the RO disappears. For $N_g(0)=0$ and $N_e(0)=1$, the decay rate approximates $\Gamma_g+\Gamma_e$ as β→∞, which can be easily checked with removing RO, and the RO occurs for $\gamma_0>0$. Since the optical transition rates are much smaller than the phonon transition, the angular frequency of RO for T≠0 has the simple form obtained by setting $\Gamma_g=\Gamma_e=0$, that is

$$\omega_{RO} = \sqrt{\Delta^2 + 4N_B\gamma_0^2} \tag{43}$$

for $N_g(0)=1$ and $N_e(0)=0$, and

$$\omega_{RO} = \sqrt{\Delta^2 + 4(1+N_B)\gamma_0^2} \tag{44}$$

for $N_g(0)=0$ and $N_e(0)=1$. In comparison with the former case, the oscillating frequency of RO for the latter case is insensitive to T.

## IV. Numerical Results and Discussions

The TRPL intensity of the system to the line feature of the state |i> can be related to the excitonic density $N_i(t)$ by $S(t) \propto |N_i(t)|^2$. Inverse Laplace transforms of Eqs. (39) and (40) give the time evolution of $N_g(t)$ and $N_e(t)$, respectively. The excessive dissimilarity of the decay time from the oscillating period of RO makes one hard to well display both features of the time decay and the period in a figure. In order to demonstrate these features without losing its generality, an unusual transition rate $\Gamma_g = 0.1 meV$ of the state |g> is used in the calculation of Fig. 5. In the figure, the ROs appear coherently with a phase separation π between $N_g$ and $N_e$, because the exciton coherently mediated by the LO phonons moves up and down between these states. The period of the oscillations for both cases of $N_g(0)=0$ and $N_e(0)=1$ (Fig. 5(a)), and $N_g(0)=1$ and $N_e(0)=0$ (Fig. 5(b)) shows to be consistent with the results calculated from Eqs. (43)-(44), the values estimated to be ~0.16ps and ~0.83ps, respectively. The increase of temperature increases its transition probability and hence results to higher oscillating amplitude and oscillating frequency. The figure also reveals the numerical results of $N_e(t)$ (dot) and $N_g(t)$ (dash-dot) with removing the RO, which well describe the time decay of the excitonic densities and will be used to extract the decay time by simple-exponential fitting.

The amplitude and frequency of the RO are remarkably dependent on the initial distributions of the excitonic population. For $N_g(0)=1$ and $N_e(0)=0$, the resonant-round trip of the exciton between |g> and |e> is stimulated and continued by repeatedly absorbing and emitting an LO phonon with the same phonon's momentum q. In this case, the amplitude and frequency are very sensitive to T and Δ (see Fig. 6), where both of them increase with increasing T, because the rise of the number of phonons around the exciton increases the probability of phonon-assisted transition. In the calculation of Fig. 6, the parameters of $\gamma_0 = 5.69 meV$ and $\Gamma_g \cong 0.355 \mu eV$ (for a wide range ~10meV of the detuning energy) estimated in Sec. II with the inter-dot distance ~4.2nm are used. At high T, an identical occupation probability in both states causes the amplitude approaches its maximum 0.5 due to the particle conservation $N_g(0)+N_e(0)=1$ for only one exciton in the system. For $N_g(0)=0$ and $N_e(0)=1$, the exciton initially occupied in the excited state moves down and up between these states through spontaneously emitting a LO phonon and, then, absorbing the phonon. The spontaneous emission of LO phonon makes the resonant-round trips happen even at T=0. In this case, the amplitude and frequency are insensitive to T and $\Delta$.

In order to discuss the decay time for these systems, a simple-exponential fitting is used to extract the decay time from the curves calculated by Eqs. (39) and (40) with removing the RO. In the limit of $\beta=0$ and/or T=0 and the exciton initially occupied in the state |g>, zero phonon decouples these states, thus, the decay rate of the exciton is solely determined by the spontaneous-emission (SE) rate $\Gamma_g$ of the state |g>. At finite T and $\beta \neq 0$, the exciton distribution dominates the decay rate which takes the form $N_{gNRO}(0^+)\Gamma_g + N_{eNRO}(0^+)\Gamma_e$. In the limit of $\beta \to \infty$ and/or T$\to \infty$, the exciton has identical probability (~0.5) to stay in both the states |g> and |e> with the decay rate $(\Gamma_g + \Gamma_e)/2$. The decay rate is one half of the SE rate (or double its corresponding decay time) for a fully-dark state $\Gamma_e = 0$, and slightly changes for $\Gamma_e = 0.9\Gamma_g$. For the excited state that is brighter than the ground state, for example $2\Gamma_g$, the decay rate increases to three half of the SE rate. All these features are revealed in Fig. 7. Because the brightness of the excited state of a QDM system vanishes for a symmetric structure or is weak for an asymmetric one, the maximal increase on its decay time at high T approaches two — it is comparable to the experimental result that shows to be slightly more than two (solid square in Fig. 8). This is different from QD systems, where the brightness of the excited state in the system is commensurate with that of the ground state. Hence, the change in the decay time (solid triangle) with respect to T is not conspicuous. The experimental data also shows a rapid decease in decay time as T>100K for both QD and QDM due to the thermal emission of the carriers out of the QD/QDM, which does not take into account in this work.

## V. Conclusion

In conclusion, the theory of exciton coupling to photons and LO phonons in QDs/QDMs was derived. Resonant-round trips of excitons between the ground and excited states mediated by LO-phonons alter the decay time and exhibit a RO. The decay time is strongly dependent on the brightness of the excited state — a dark state results to enhance the decay time (the case of QDMs), and a bright state to reduce the decay time (the case of QDs) — and the detuning energy between these states. A three-level system mediated by one-phonon process gives the maximum enhancement of the decay time by a factor of two. In strong coupling regime, a further consideration of multiple levels with multiple-phonon processes has to be done, since the manifold loops of polaron resonance mediated by multiple phonons could also affect the decay time. For example, there are two types of polaron resonance including two-phonon process as schematically plotted in Fig. 9. In the first one (Fig. 9(a)), the exciton takes more time than three-level system in resonant-round trips between the states $|g\rangle$, $|e1\rangle$, and $|e2\rangle$. At high T and/or strong coupling region, the identical probability (~1/3) of the exciton distribution in these states reduces the decay rate to one-third that of zero-T. While the second (Fig. 9(b)) decreases the decay rate nothing less than the three-level system, because the exciton has the probability of ~1/2 to stay in the state $|g\rangle$ and only the probability of ~1/4 in the states $|e1\rangle$ and $|e2\rangle$. In conclusion, the increase of the excitonic decay time in QDs/QDMs is resulted from the total effect of multiple weak-bright excited states or from a major state provided with a weak-optical transition and a small detuning energy. Moreover, the population-dependent amplitude and frequency of RO provide a detectable signature to the information stored in QD/QDM systems for a wide range of temperature. This is useful in the application of quantum information processing.

**Acknowledgement**

This work was supported by the National Science Council of the Republic of China, Taiwan, under Contract No. NSC95-2623-7-151-003-D, and the Ministry of Education of R. O. C.

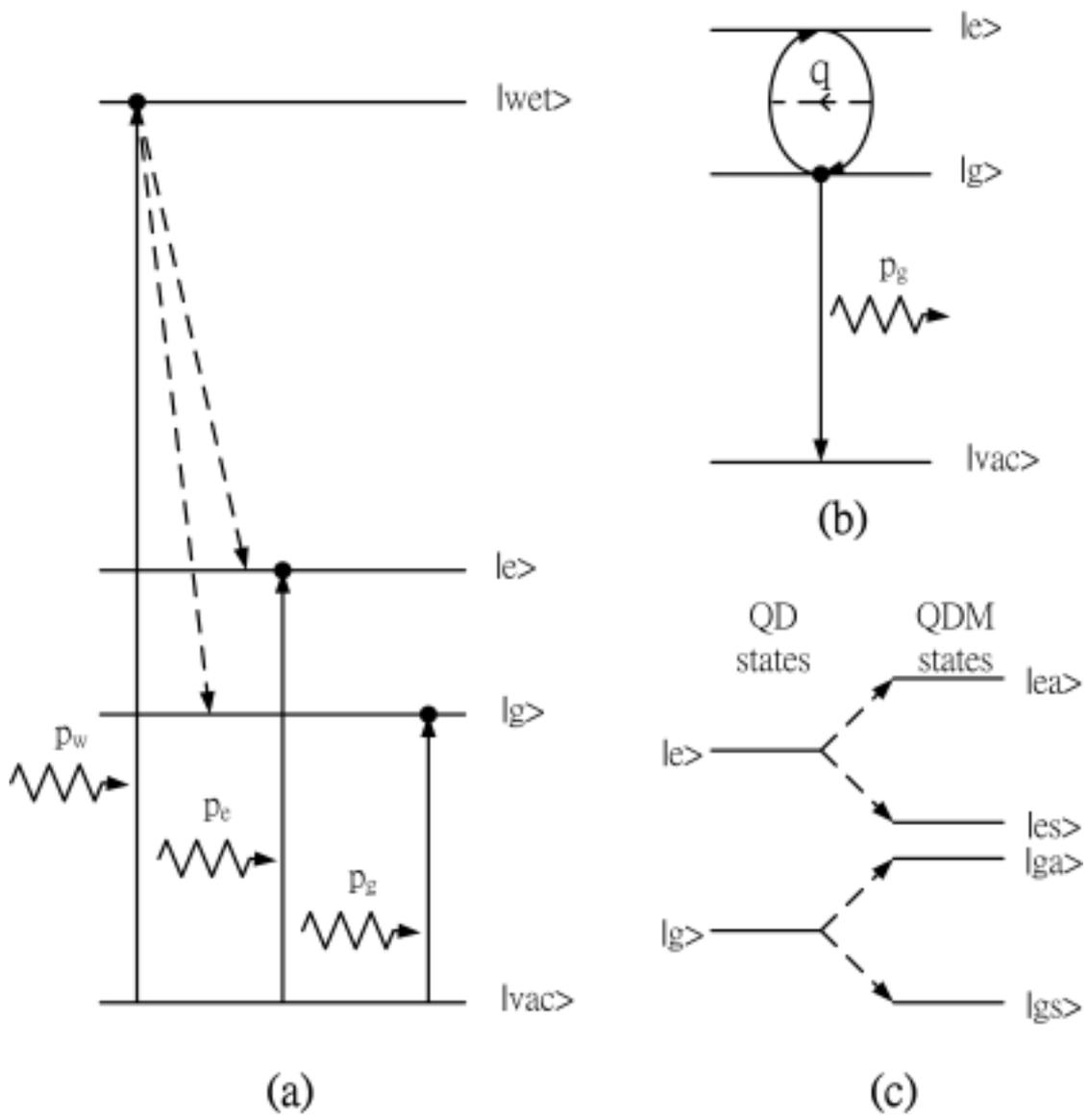

Fig. 1. (a) Schematic plot of the excitation processes. (b) The plot of a three-level excitonic system for QD or QDM, where solid-wiggled (dashed) lines denote the photon (phonon) absorption/emission. (c) Schematic plot of the splitting of the QD states into symmetry (|gs> and |es>, bright) states and anti-symmetry (|ga> and |ea>, dark) states for QDM.

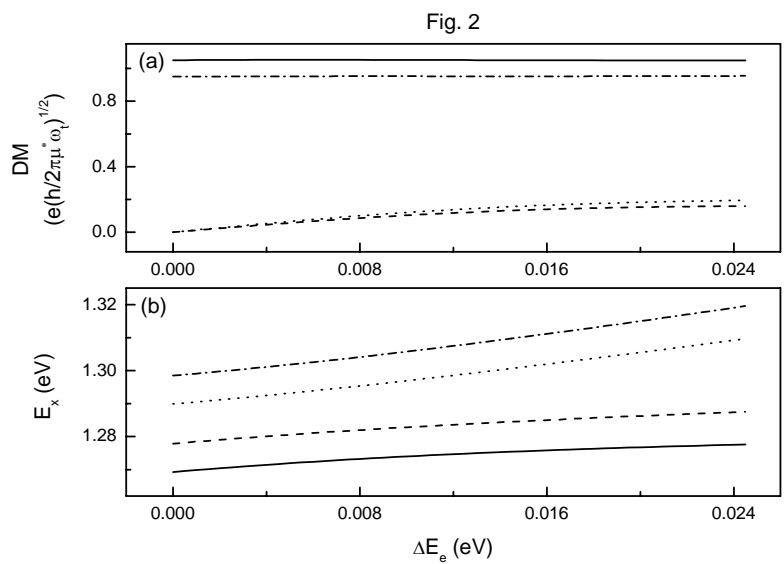

Fig. 2. Calculated results of (a) the DMs and (b) the exciton energies with respect to $\Delta E_e$ for the states $|S1\rangle$ (solid), $|S2\rangle$ (dash), $|S3\rangle$ (dot), and $|S4\rangle$ (dash-dot) in the distance of 5nm.

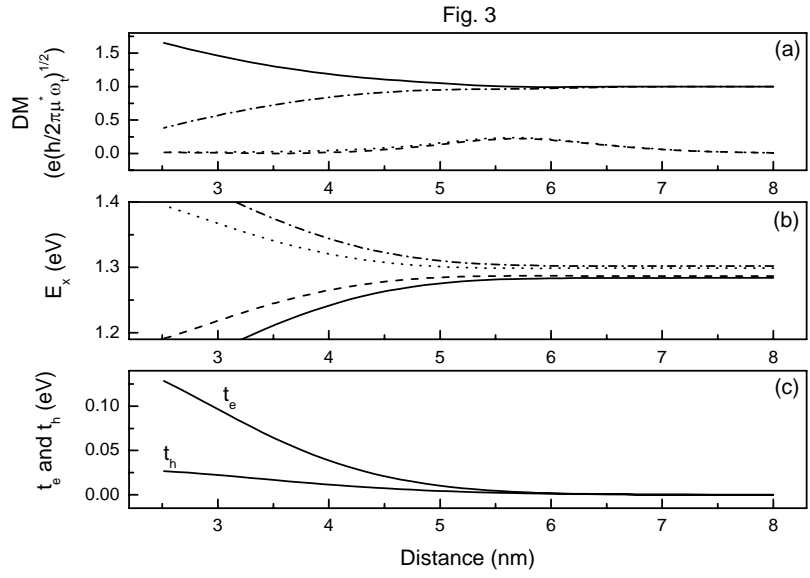

Fig. 3. Calculated results of (a) DMs, (b) tunneling strengths, and (c) exciton energies with respect to the distance of the dots for the states |S1⟩ (solid), |S2⟩ (dash), |S3⟩ (dot), and |S4⟩ (dash-dot) in $\Delta E_e = 15$ meV.

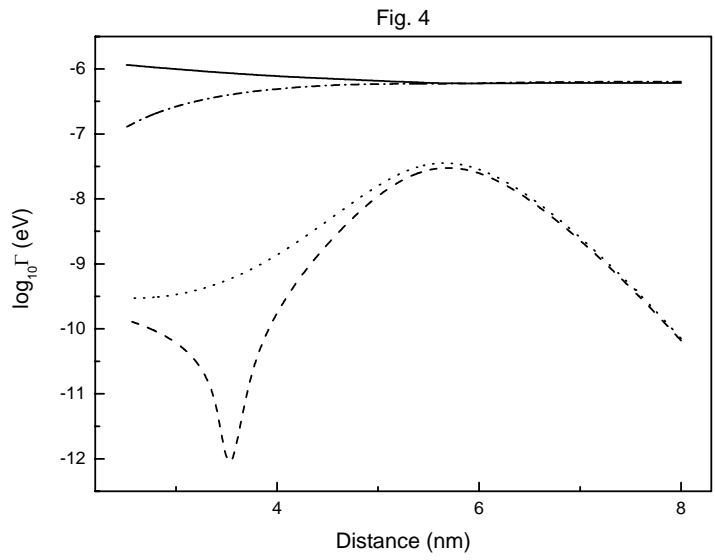

Fig. 4. Calculated results of the transition rates with respect to the distance of the dots for the states |S1> (solid), |S2> (dash), |S3> (dot), and |S4> (dash-dot) in $\Delta E_e$=15meV.

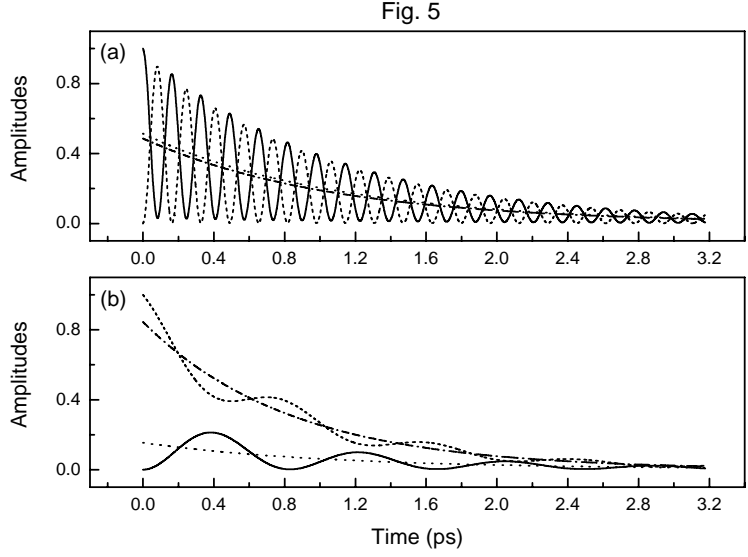

Fig. 5. Time evolution of the excitonic densities in the ground state (dash) and the excited state (solid) with the parameters of $\omega_{LO} = 36 meV$, $\Gamma_g = 0.1 meV$, $\Gamma_e = 0.5\Gamma_g$, $\gamma_0 = 3 meV$, $\Delta = 1 meV$, and the initial conditions of (a) $N_g(0)=0$ and $N_e(0)=1$, and (b) $N_g(0)=1$ and $N_e(0)=0$ at T=95K. The calculated angular frequencies for the conditions of (a) and (b) are, respectively, $\sim 6.12 meV/\hbar$ and $\sim 1.2 meV/\hbar$. The dot (dash-dot) line displays the density of $N_e(t)$ ($N_g(t)$) with removing RO for $\beta = 3$.

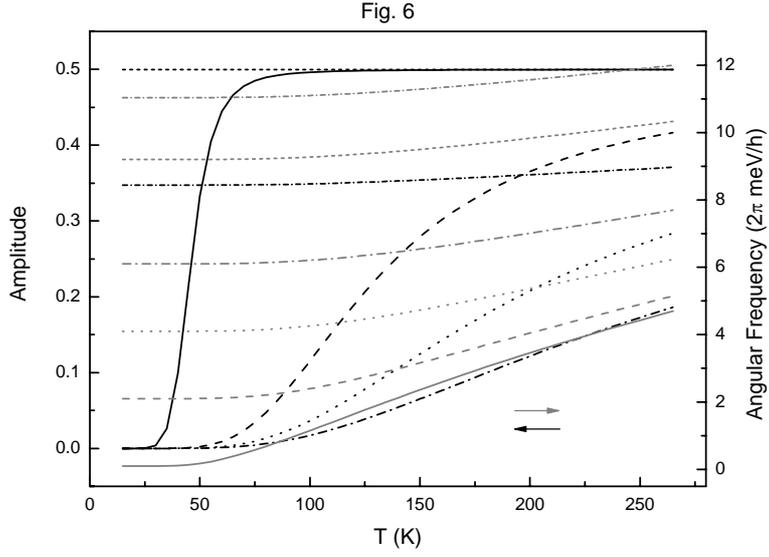

Fig. 6. The oscillating frequency (gray) and oscillating amplitude (black) of the RO with the parameters of $\Gamma_g = 0.355\mu eV$, $\Gamma_e = 0$, $\gamma_0 = 4.6 meV$ for Δ=0.1meV (solid), 2.1meV (dash), 4.1meV (dot), and 6.1meV (dash-dot) with the initial conditions of $N_g(0)=1$ and $N_e(0)=0$, and for Δ=0.1meV (short dash) and 6.1meV (short dash-dot) with the conditions of $N_g(0)=0$ and $N_e(0)=1$.

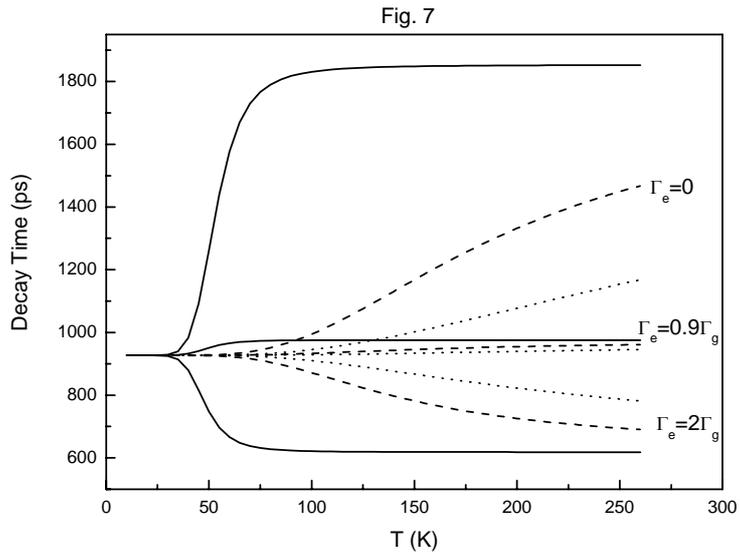

Fig. 7. The changes in decay time relative to that of T=0 as a function of T and $\Gamma_e$ for $\beta = 1.12$ (dash), $\beta = 2.19$ (dot), and $\beta = 46$ (solid) with the initial conditions of $N_g(0)=1$ and $N_e(0)=0$.

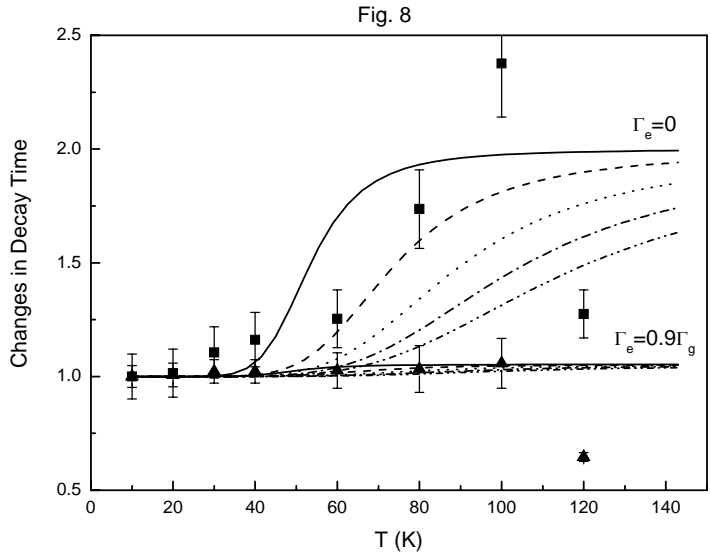

Fig. 8. The changes in decay time relative to that of T=0 as a function of T and $\Gamma_e$ for $\beta = 5.11$ (dash-dot-dot), $\beta = 6.57$ (dash-dot), $\beta = 9.2$ (dot), $\beta = 15.3$ (dash), and $\beta = 46$ (solid), with the initial conditions of $N_g(0)=1$ and $N_e(0)=0$. The solid square (triangle) denotes the experimental results of InGaAs QDMs [6] (InGaAs QDs [6]).

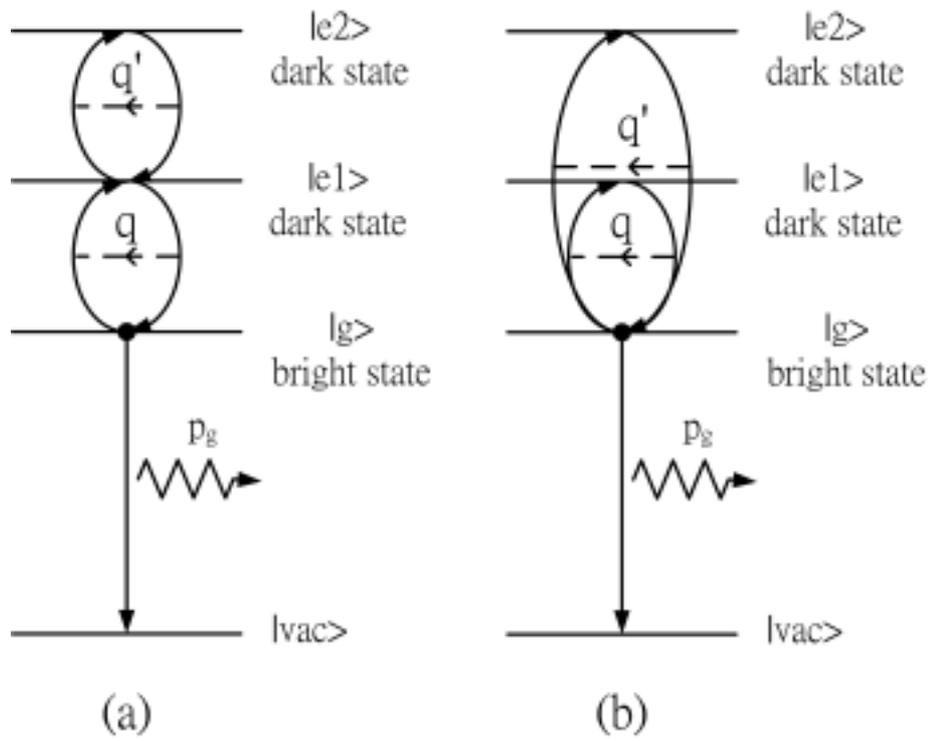

Fig. 8. Schematic plot of a four-level system.